\documentclass{article}
\usepackage[latin1]{inputenc}
\usepackage{slashed}
\usepackage{amsmath}
\usepackage{textcomp}
\usepackage{amssymb}
\usepackage{amsfonts}
\usepackage{indentfirst}
\usepackage{color}
\usepackage{hyperref}
\usepackage{appendix}
\usepackage[top=2cm,bottom=2cm,left=3cm,right=2cm]{geometry}
\newcommand{\nin}{\noindent}
\newcommand{\be}{\begin{equation}}
\newcommand{\ee}{\end{equation}}
\newcommand{\bea}{\begin{eqnarray}}
\newcommand{\eea}{\end{eqnarray}}

\newcommand{\nn}{\nonumber\\}

\usepackage{graphicx,graphics}
\graphicspath{{figures/}}

\begin{document}

\hfill{KCL-PH-TH/2022-19}

\begin{center}
{\Large{\bf Symmetry restoration, Tunnelling and the Null Energy Condition}}

\vspace{1cm}

{\bf Jean Alexandre}\\
Theoretical Particle Physics and Cosmology, King's College London, UK

\vspace{0.3cm}

{\bf Janos Polonyi}\\
Theory Group, CNRS-IPHC, University of Strasbourg, France
\end{center}


\vspace{1cm}

\begin{abstract}
    A finite volume allows tunnelling between degenerate vacua in Quantum Field Theory, and  
    leads to remarkable energetic features, arising from the competition of 
    different saddle points in the partition function. We describe this competition for finite temperature at equilibrium, taking into account both static and 
    (Euclidean) time-dependent saddle points. 
    The effective theory for the homogeneous order parameter yields a non-extensive vacuum energy at low temperatures,
    implying a dynamical violation of the Null Energy Condition.
\end{abstract}

\vspace{1cm}

\section{Introduction}

Spontaneous symmetry breaking (SSB) is an important and peculiar feature of many body systems. Its importance stretches from solid state 
physics to the Standard Model and its peculiarity is that it is not truly realised. In fact, SSB is valid in 
strictly infinite systems, a condition which one never meets in physics. However the order of magnitude of Avogadro's number and the 
ratio of the macroscopic and microscopic characteristic scales make SSB an excellent approximation in most cases. 
This work shows how fundamental energetic features are modified in finite systems, where SSB is not a good approximation.
More specifically, the Null Energy Condition (NEC \cite{NECreview}) can be violated, 
as shown in \cite{AC,AP,AB}, and
the present article describes the finite-temperature equilibrium state obtained from tunnelling between two vacua.\\

In the thermodynamical limit, involving an infinite volume, a first order phase transition is characterised by the Maxwell cut, signalling the degeneracy of 
the ground state of a system. The corresponding flat effective potential is consistent with convexity, obtained as a consequence of the competition of different 
key configurations in the partition function \cite{convexity}, the saddle points of the model. 
For finite systems though, symmetry is restored by tunnelling, and the effective potential obtained from the interplay of different saddle points has a non-trivial dependence
on the volume. This featured was noticed in \cite{AT}, where the one-particle-irreducible (1PI) effective potential is calculated for an $O(N)$-symmetric scalar field,
taking into account homogeneous saddle points. In \cite{AT} fluctuation factors are ignored in the semi-classical approximation for the partition function, 
but taking these factors into account leads to identical conclusions \cite{AB}. 
In these works, the Maxwell cut is recovered in the limit where the spacetime volume goes to infinity. \\

We briefly comment here on the Wilsonian running potential, obtained from exact functional renormalisation group studies \cite{exact}. 
While this potential is not necessarily convex along the renormalisation trajectory, it approaches a convex effective potential at the 
infrared end point \cite{Wilsonian}. But the Wilsonian effective potential is identical to the 1PI effective 
potential at the IR end point only for infinite spacetime volumes, whereas the present study focuses on finite volumes.\\

NEC violation is known in Quantum Field Theory, with the typical example of the Casimir effect (see \cite{Bordag} for a review).
This effect has been used in the context of Early Cosmology to induce a spacetime expansion \cite{ZS}, where NEC violation is obtained 
from a massless scalar field in a 3-torus. The Casimir effect is not only suppressed by some inverse power of the volume though, 
but it is also exponentially suppressed by the scalar field mass, and we neglect it in the present work.
In the case we study here - a massive scalar field in the presence of different vacua - NEC violation  
is expected by recalling the variational method of Quantum Mechanics, 
where the ground state energy is lowered by taking into account the mixing of the degenerate potential wells \cite{Kleinert}. 
In the present work, the NEC violation is a consequence of a non-trivial volume dependence of the effective action $S_{eff}$,
hence of non-extensive thermodynamical potentials.
The energy density $\rho$ and the pressure $p$ in the vacuum are obtained from the free energy $F=VU_{eff}(0)=TS_{eff}(0)$, and
\bea
\rho+p&=&\frac{1}{V}\left(F-T\frac{\partial F}{\partial T}\right)-\frac{\partial F}{\partial V}\\
&=&-T\frac{\partial U_{eff}(0)}{\partial T}-V\frac{\partial U_{eff}(0)}{\partial V}~.\nonumber
\eea
The first term in the second line is the usual one, arising from temperature-driven quantum fluctuations, and is positive. 
The non-trivial second term arises from tunnelling and is a consequence of the non-extensive feature of the effective action $S_{eff}(0)$.
Note that, if the partition function is based on one saddle point only, then the effective action is extensive
and this second term vanishes. In the situation of several saddle points though, we show in this article that 
this term is negative, and quantum fluctuations dominates over thermal fluctuations at low enough temperature.
In our work, NEC violation is therefore a finite volume effect. 
As shown in \cite{AC}, the corresponding mechanism could be relevant for the generation a cosmological bounce \cite{cosmobounce}, without the need for modified gravity or exotic matter.\\

We stress here that this work is not related to the Kibble-Zurek mechanism \cite{KZ}. The latter necessitated a high temperature in order to 
restore symmetry, whereas the mechanism we present here is valid for any temperature, including zero-temperature, where SSB would indeed occur in an infinite volume. 
This is detailed in Section \ref{Secversus}, where we also explain that we do not take into account bubbles of different vacua, because we work at finite volume.
In Section \ref{Secsaddle} we describe the different saddle points relevant here, which are the static ones and the time-dependent instanton/anti-instantons pairs,
hence going further than the work \cite{A}, which take into account static saddle points only. 
The contribution of these saddle points to the semi-classical expansion is discussed in Section \ref{semexps}.
We study then the intermediate-temperature regime in Section \ref{SecintermT},  where the Euclidean time is not large enough to allow the formation of instanton/anti-instanton pairs, and only the static saddle points contribute to the partition function. 
Although the effective potential has a non-extensive field dependence, the vacuum energy is intensive, such that
the NEC is satisfied in the vacuum.
Section \ref{SecsmallT} describes the low temperature regime, dominated by a dilute gas of instanton/anti-instantons, 
which allow for a non-trivial volume-dependence of the vacuum energy. As a result, for a fixed volume, we show that we can always find 
a temperature small enough for the NEC to be violated.

\section{Tunnelling versus Thermal Symmetry Restoration}
\label{Secversus}

We consider the bare Euclidean action for a scalar field $\phi(t,\vec x)$ at temperature $T=1/\beta$ and three dimensional spatial volume $V=L^3$
\be\label{Sbare}
S_{bare}[\phi]=\int_0^\beta dt\int_Vd^3x\left[\frac12\partial_\mu\phi\partial_\mu\phi+\frac{m_b^2}2\phi^2+\frac{\lambda_b}{4!}\phi^4\right]~.
\ee
The parameters $m_b^2,\lambda_b$ of the bare theory are chosen in such a manner that the vacuum displays the spontaneously broken symmetry $\phi\to-\phi$. 
Therefore the order parameter, the space-time average of the field, develops non-vanishing expectation value in the thermodynamical limit $V\to\infty$, 
and for temperatures below the critical temperature which appears in the
Kibble-Zurek mechanism
\be
T<T_c=2v_0~,
\ee
where $v_0$ is the dressed vacuum expectation value (vev) for vanishing temperature.

Exact SSB may occur only in a strictly infinite system which is a formal construct since one always encounters finite objects in physics. 
A finite physical system is said to display broken symmetry if SSB becomes a better and better approximation as the volume increases without limit. 
Hence SSB is a particular asymptotic finite size scaling law. 

Let us start with an infinite system in a symmetry broken vacuum where the SSB is exact and distinguish two symmetry restoration mechanisms. 
The symmetry is restored thermally if thermal fluctuations have sufficient energy to spread the density matrix over different degenerate vacua. 
Such a symmetry restoration mechanism is active even in the strictly infinite volume case, and happens for $T>T_c$. 
The alternative symmetry restoration process, tunneling under the potential barrier separating the degenerate vacua, operates only in finite volume. 
As was observed in \cite{AC,AP,AB}, the non-extensive feature of the thermodynamical potential due to such a symmetry restoration in finite systems leads to NEC violation. 
Extensivity of a quantity is defined for lengths large compared to the characteristic length of a system though, such that
our main interest in this work is to find the volume dependence of the thermodynamical potential for large volume.

The improvement of the SSB approximation with increasing volume is related to a tunneling time $\tau_t(V)$ which increases exponentially with the volume $V$. 
In addition, the tunnelling dynamics depends on the temperature $T=1/\beta$, in such a way that the thermodynamical properties of the system are controlled by the dimensionless 
ratio $\beta/\tau_t(V)$, c.f. eq. (\ref{Nbar}) below. Tunneling ceases to be active for $V\gg V_\beta$ where $V_\beta$ is defined by $\beta=\tau_t(V_\beta)$.
As shown in Section \ref{SecsmallT}, the latter identity leads to a temperature $T(V)$ which decreases exponentially with the volume, and
below which extensive properties are recovered.\\

We are interested in the effective dynamics of the order parameter because it may have an important role in violating NEC. 
This effective dynamics is defined by the Wilsonian action $S[\phi]$ for the spatially homogeneous field component, $\phi(t)$. 
It is important to distinguish the full dynamics of the field $\phi(t,\vec x)$ and the effective dynamics of the order parameter $\phi(t)$: 
The spontaneous breakdown or the restoration of the symmetry is driven by the former, whereas the latter only provides a diagnostic of the status of symmetry.

To probe the possible non-extensive feature of the thermodynamical potentials in the symmetry broken phase we need the partition function of the full theory as the function of the volume and the temperature. 
The status of symmetry impacts the partition function, with an effect encoded by the effective potential for $\phi(t)$. 
To obtain the effective potential in the one-loop saddle point expansion, we assume the following action to describe the effective dynamics
(see Appendix \ref{wapp} for details)
\be\label{wes}
S[\phi]=V\int_0^\beta dt\left(\frac12(\partial_0\phi)^2+\frac\lambda{24}(\phi^2-v^2)^2+j\phi\right)~,
\ee
which features a double well structure. It is argued below that the qualitative aspects of our results remain valid beyond this particular form of the action.\\

We end this section with a comment on the choice of a homogeneous order parameter $\phi(t)$. 
For a non-vanishing source $j$, the vacuum degeneracy is shifted, 
and one should in principle consider coexisiting domains of true and false vacuum.
Following \cite{CC}, it is reasonable to assume spherical domains of radius $R$, whose actions are of the form
\be
S_{bubble}=-ajR^3+ bR^2~, 
\ee
where $a>0$ and $b>0$. In the previous expression, the volume contribution is proportional to the energy difference between the vacuua, 
and competes with the surface term arising from surface tension. The resulting critical bubble radius is
\be
R_{cr}=\frac{2b}{3aj}~,
\ee
which diverges for the ground state of the system, obtained for $j\to0$ because of symmetry restoration. As a consequence, finite volumes on which the
present study is based do not allow the formation of bubbles, as long as one focuses on the vicinity of the true ground state. This justifies the study of
homogeneous and time-dependent saddle points in this work.

\section{Saddle points for homogeneous dynamics}
\label{Secsaddle}
We introduce the dimensionless variables
\be
\tau=\omega t~~~~,~~~~\varphi\equiv\sqrt\frac{\lambda}{6}~\frac{\phi}{\omega}~~~~,~~~~
~k\equiv\sqrt\frac{\lambda}{6}~\frac{j}{\omega^3}~,
\ee
with $\omega\equiv v\sqrt{\lambda/6}$, we are lead to the bare action 
\be\label{S}
S=B\int_{-\omega\beta/2}^{\omega\beta/2}d\tau\left((\varphi')^2+\frac{1}{2}(\varphi^2-1)^2+2~k\varphi\right)~,
\ee
where a prime denotes a derivative with respect to $\tau$ and
\be\label{B}
B\equiv\frac{3}{\lambda}(L\omega)^3~.
\ee

The equation of motion obtained from the action (\ref{S}) is
\be\label{equamot}
\varphi''+\varphi=\varphi^3+~k~.
\ee
Below we study the different solutions of this equation, which are relevant to finite-temperature field theory, where periodic boundary conditions are imposed on field configurations. We note that there are other solutions of the equation of motion, which are relevant in the context of false vacuum decay, as the "shot" \cite{Andreassen}, which doesn't satisfy periodic boundary conditions though, and is not considered in the present study.

The partition function is an analytic function of $k$ for finite volume $V<\infty$ and finite temperature $\beta<\infty$. 
As a consequence, the formal symmetry $\phi\to-\phi$ and $k\to-k$ assures that the partition function depends on $k^2$, 
although individual saddle points are not necessarily even in the source.

In what follows we consider an infinitesimal source $k$, since we focus on the ground state of the system: Because of tunnelling,
this ground state corresponds to a vanishing classical field, which in finite volume maps to a vanishing source though the Legendre transform.

\subsection{Static saddle points}

Static vacuua satisfy 
\be
\varphi^3-\varphi+k=0~,
\ee
and the number of solutions depends on the dimensionless source $k$: if we define $k_c\equiv2/3\sqrt3\simeq 0.385$,
we have the following two regimes:\\
$\bullet$ For $|k|>~k_c$, the model has only one (real) static vacuum, which is
\be\label{varphi0s}
\varphi_{s0}=-\mbox{sign}(k)\frac{2}{\sqrt3}\cosh\left(\frac{1}{3}\cosh^{-1}(|k|/k_c)\right)~.
\ee
$\bullet$ For $|k|<~k_c$ the model has two static vacuua, which are 
\be\label{phi1phi2}
\varphi_{s1}(k)=\varphi_s(k)~~~~\mbox{and}~~~~\varphi_{s2}(k)=-\varphi_s(-k)
\ee
where
\be\label{phis}
\varphi_s(k)=\frac{2}{\sqrt3}\cos\left(\frac{\pi}{3}-\frac{1}{3}\cos^{-1}(k/k_c)\right)~,
\ee
and which corresponds to the regime we focus on in this article.
The actions for these saddle points are
\be\label{S1S2}
S_1\equiv S[\varphi_{s1}]=\Sigma(k)~~~~,~~~~S_2\equiv S[\varphi_{s2}]=\Sigma(-k)~,
\ee
where
\be\label{Sigma}
\Sigma(k)=B\omega\beta\left(2k-\frac{k^2}{2}-\frac{k^3}{4}-\frac{k^4}{4}-\frac{21}{64}k^5-\frac{k^6}{2}+{\cal O}(k^7)\right)~,
\ee
where we note that they differ only by the sign of the odd powers of the source $k$. This will be important to recover the appropriate parity of the effective potential.

\subsection{Time-dependent saddle points}

The Euclidean time-dependence of a saddle point can be found from the Minkowski time-dependence after the sign flip $U(\phi)\to-U(\phi)$. 
There are many periodic solutions of eq.(\ref{equamot}), whose qualitative features can better be understood by 
following their period length as the function of the maximal value, $\varphi_m=\max_\tau(\varphi(\tau))$. 
The solutions with the shortest period $2\pi/\omega$ are harmonic oscillations around $\varphi=0$ with infinitesimal $\varphi_m$. 
Hence there are no time-dependent saddle points at high temperature, $\beta<2\pi/\omega$. The increase of $\varphi_m$ leads longer periods, 
since the quartic part of the potential weakens the restoring force to the equilibrium position $\varphi=0$. 
As $\varphi_m$ approaches $\max\{\varphi_{s1},\varphi_{s2}\}$, the trajectory spends most of the time around one of these static saddle points and the period length diverges. 
The action is a decreasing function of $\varphi_m$ hence our interest lies mainly in low temperature time-dependent saddle points. The periodic saddle points are  instanton/anti-instanton pairs, and each instanton or anti-instanton has the approximate width $1/\omega$. The number of pairs allowed
in the Euclidean total time $\beta$ is not a continuous function of $\beta$, and their maximum number is 
\be\label{Nbeta}
N_\beta\sim\frac{\omega\beta}{2\pi}~.
\ee
We are thus led to two saddle point regimes, that we will study separately:
\begin{itemize}
    \item The intermediate-temperature regime $\beta_c<\beta<2\pi/\omega$, where only static saddle point are present. 
    This is a wide temperature interval if $\omega\beta_c\ll2\pi$;
    \item The low-temperature regime $2\pi/\omega\ll\beta$, where instanton/anti-instanton pairs can develop.
\end{itemize}

\subsubsection{Instanton/anti-instanton pair}

\begin{figure}[h!]
    \centering
    \includegraphics{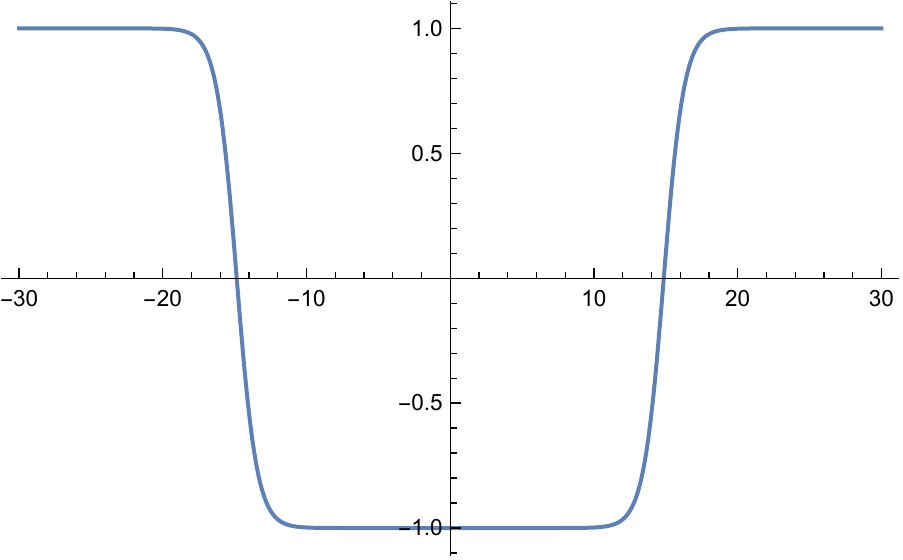}
    \caption{An instanton/anti-instanton pair within the period $\omega\beta=60$, obtained with $\varphi_0(0)\simeq-0.999999997$.
    In the absence of source, the configuration spends the same time close to both static saddle points.}
    \label{phi01}
\end{figure}

For large but finite $\omega\beta$, a instanton/anti-instanton pair spends most of the Euclidean time close to 
the static saddle points $\pm v$ when $k=0$. Figure \ref{phi01} shows such a pair, which is exactly symmetric and spends the same time close to 
the two vacua. This pair can be approximated by
\be\label{approxphi01}
\varphi_0\simeq\tanh\left(\frac{\tau-\frac{\omega\beta}4}{\sqrt2}\right)\tanh\left(\frac{\tau+\frac{\omega\beta}4}{\sqrt2}\right)~,
\ee
since, for both signs of $\tau$, one of the hyperbolic tangent factors is very close to $\pm1$, and the other hyperbolic tangent factor satisfies the equation of motion. 
Apart from a time approximately equal to $2/\omega$, the saddle point is exponentially close to the vacuum hence the action for the configuration arises mainly from 
the jumps only and can be approximated by
\be\label{S0}
S_0=\frac{8\sqrt2}{3}~B~.
\ee

If one introduces an infinitesimal source $k\ne0$, the action for one instanton/anti-instanton pair is still dominated half of the Euclidean time $\beta$ 
by the action of the saddle point $\phi_{s1}$, and the other half by the action of the saddle point $\phi_{s2}$. 
The subdominant part of the action arises from the jumps of the instanton and anti-instanton, and the pair action is 
\be\label{SB1}
S_p(k)\simeq S_0+\frac{1}{2}\Sigma(k)+\frac{1}{2}\Sigma(-k)~,
\ee
where $S_0$ and $\Sigma$ are given by eq.(\ref{S0}) and eq.(\ref{Sigma}) respectively.

\subsubsection{Anti-ferromagnetic saddle point configurations}

The regime $\omega\beta\gg1$ allows multiple instanton/anti-instanton pairs exact saddle points, which are regularly distributed in the Euclidean period $\beta$. 
The corresponding action is even in $k$ and has the generic form
\be
S^{multi}_p(k)=\sum_{l=0}^\infty s_{2l}k^{2l}~.
\ee
If the instantons and anti-instantons are far enough from each other though, the configuration still spends most of the time close to one saddle point, and
the total contribution from the $n$ pairs is approximately $nS_0$. The action is thus, for an infinitesimal source,
\be\label{SBn1}
S_p^{(n)}(k)\simeq nS_0+\frac{1}{2}\Sigma(k)+\frac{1}{2}\Sigma(-k)~,\\
\ee
and a discussion on dilute gas of instanton/ant-instanton is given in the next section.

\section{Semi-classical approximation}\label{semexps}

In the presence of multiple saddle points $\phi_n$ and if one doesn't allow SSB, the partition function can be approximated by the sum of path integrals,
each integrating fluctuations perturbatively around $\phi_n$
\be
Z[j]\simeq\sum_i\int{\cal D}[\xi]\exp(-S[\phi_i+\xi])~,
\ee
where the action $S[\phi_i+\xi]$ involves the source term. This semi-classical approximation becomes better as the spacetime volume $V\beta$ is large, 
since fluctuations around each saddle point are suppressed exponentially and do not communicate. We have thus
\be
\int{\cal D}[\xi]\exp(-S[\phi_i+\xi])\simeq F_i e^{-S[\phi_i]}~,
\ee
where the factors $F_i$ take into account fluctuations determinants for time-dependent modes,
but also the eventual zero mode corresponding to the translational invariance of instantons/anti-instantons, as explained below.

\subsection{Static saddle points}

The fluctuation factor arising from time-dependent fluctuations over a static saddle point is calculated in Appendix \ref{1loopApp}, and we have 
\be
F_1=F_\beta(k)~~~~\mbox{and}~~~~F_2=F_\beta(-k)~, 
\ee
where
\be\label{Fofk}
F_\beta(k)=\exp\Big(-B\ln\sinh\big((\omega\beta/2)\sqrt{3\varphi_s^2(k)-1}\big)\Big)~,
\ee
and $\varphi_s(k)$ is given in eq.(\ref{phis}). Note that, for $\omega\beta\gtrsim2$ we have 
\be\label{asymptF}
F_\beta(k)\simeq\exp\left(-\frac{B\omega\beta}{2}\sqrt{3\varphi_s^2(k)-1}\right)~,
\ee
which will be used further on.

\subsection{Instanton/anti-instanton pair}

Following \cite{Kleinert}, the fluctuation factor for an instanton or an anti-instanton spending time $\beta_1$ close to one static saddle point 
and the time $\beta_2$ close to the other static saddle point, with $\beta_1+\beta_2\simeq\beta$ is
\be
F_{inst}\simeq \sqrt\frac{S_0}{2}F_{\beta_1}(k)F_{\beta_2}(-k)\omega\beta~,
\ee
where the factor $\sqrt{S_0/2}~\omega\beta$ arises from the zero mode corresponding to the translational invariance of 
the configuration over the total length $\omega\beta$. 

For an instanton/anti-instanton pair, with negligible correlation for $\beta\omega\gg1$, the overall saddle point spends some time $\beta_1$ close to one static saddle point, 
then some time $\beta_2$ close to the other static saddle point, and back close to the first one for some time $\beta_3$, with $\beta_1+\beta_2+\beta_3\simeq\beta$. 
The centre of the instanton can be placed freely over a time interval of length $\omega\beta$ and the centre of the anti-instanton can be placed in the remaining time, 
leading to the zero-mode factor 
\be
\sqrt\frac{S_0}{2} \int_{-\omega\beta/2}^{\omega\beta/2}d\tau_1~\sqrt\frac{S_0}{2}\int_{\tau_1}^{\omega\beta/2}d\tau_2
=\frac{1}{4}S_0(\omega\beta)^2~.
\ee
On average $\beta_1+\beta_3=\beta_2=\beta/2$ such that the total fluctuation factor for an instanton/anti-instanton pair is
\be
F^{(1)}_p\simeq \frac{S_0}{4}(\omega\beta)^2 F_{\beta_1}(k)F_{\beta_2}(-k)F_{\beta_3}(k)~,
\ee
and, given the asymptotic form (\ref{asymptF}), 
\be\label{F1bounce}
F^{(1)}_p\simeq \frac{S_0}{4}(\omega\beta)^2 F_{\beta/2}(k)F_{\beta/2}(-k)~.
\ee

\subsection{Instanton/anti-instanton gas}

The exact solutions of the equation of motion (\ref{equamot}) involving several pairs of instanton/anti-instanton form an "anti-ferromagnetic crystal",
with a rigid structure and therefore with a small configuration-space measure in the partition function. The approximate saddle points
made of weakly coupled instanton/anti-instanton, where the later are free to be moved around in time, have a huge degeneracy arising form translational zero modes, 
without increasing much the action of the total configuration.
Because of this degeneracy, at low enough temperature they dominate over the exact crystal-structured saddle points.
The zero mode for $n$ instanton/anti-instanton pairs is then
\be\label{degeneracy1}
\sqrt\frac{S_0}{2}\int_{-\omega\beta/2}^{\omega\beta/2}d\tau_{2n}~\sqrt\frac{S_0}{2}\int^{\omega\beta/2}_{\tau_{2n}}d\tau_{2n-1}
\cdots\sqrt\frac{S_0}{2}\int^{-\omega\beta/2}_{\tau_2}d\tau_1 
=\frac{S_0^n}{2^n}\frac{(\omega\beta)^{2n}}{(2n)!}~.
\ee
On average, the anti-ferromagnetic saddle point spends the overall time $\beta/2$ close to both static saddle point and, 
using the asymptotic form (\ref{asymptF}), we obtain for the total fluctuation factor for $n$ pairs 
\be\label{Fnbounce}
F^{(n)}_p\simeq \frac{S_0^n}{2^n}\frac{(\omega\beta)^{2n}}{(2n)!} F_{\beta/2}(k)F_{\beta/2}(-k)~.
\ee

\subsection{Comments on the bounce saddle point}

In the present study we focus on an infinintesimal source $k$, which is why the relevant time-dependent saddle points consist in pairs of instanton and anti-instantons.
The latter configurations do not generate negative fluctuation modes, which is why the above fluctuation factors are all real.

\begin{figure}[h!]
    \centering
    \includegraphics{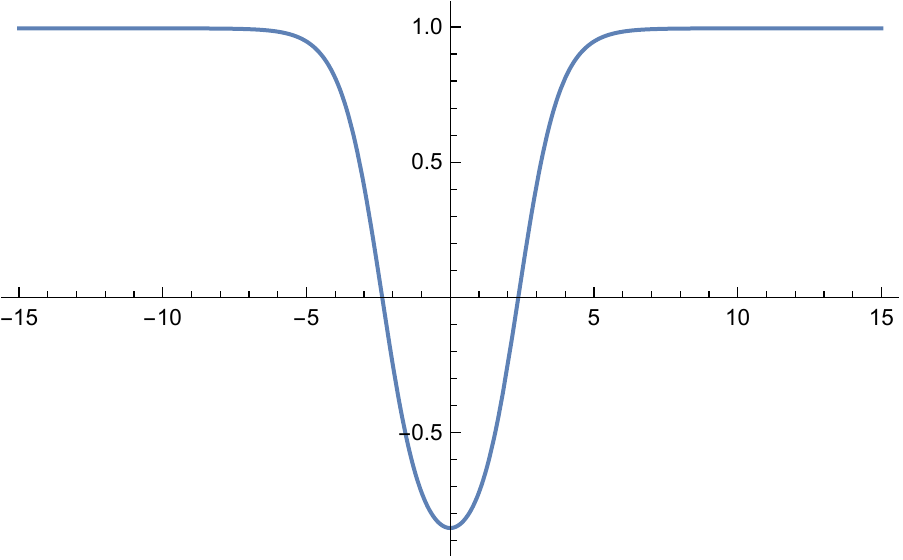}
    \caption{A single-bounce configuration within the period $\omega\beta=30$, obtained with the source $k=0.01$ and $\varphi_B(0)\simeq-0.853183042684$.
    A finite source breaks the symmetry between the vacua, and the bounce cannot be decomposed as a dilute instanton/anti-instanton pair.}
    \label{bounce}
\end{figure}

If one considers a larger source though, the time-dependent saddle point cannot be decomposed as dilute pairs of instantons and anti-instantons, and the basic building block 
of the dominant configurations is a bounce instead, as shown in Fig.\ref{bounce}. Such a bounce does induce an imaginary fluctuation determinant, arising from one negative mode
which should be treated appropriately with an analytical continuation, for the integration over quadratic fluctuations \cite{Kleinert}. 
This feature is used to determine the decay rate of a false vacuum in $O(4)$-symmetric Euclidean coordinates \cite{CC}, 
and does not appear in our study, which focused on the vicinity of $k=0$.

Should we include higher orders in the classical field for the effective potential, we would need to take a larger source, and thus consider the bounce imaginary fluctuation factor. The partition function is real though, and so must be the effective potential which describes an equilibrium state,
such that no imaginary part should appear in the effective dynamics. 
As explained in \cite{Andreassen}, when integrating quadratic fluctuation, the analytic continuation to avoid the bounce negative mode requires a contour of integration 
which goes through the other saddle points. The resulting closed contour necessarily involves additional imaginary parts, which exactly cancel the imaginary part 
arising from the bounce, indeed leading to real effective dynamics.

\section{Intermediate-temperature regime}
\label{SecintermT}

We consider here the intermediate temperature regime $\beta_c<\beta\lesssim2\pi/\omega$, dominated by static saddle points.

\subsection{Effective potential}

Although in this work we ignore spatial fluctuations, we rescale the partition function $Z_\beta[k]$ by a source-independent term $R_\beta$, 
in order to reproduce the usual ground state (3+1)-thermal fluctuations contribution, at each static saddle point. 
This contribution is
\be\label{U0}
U_0=\frac{T^4}{2\pi^2}I(\sqrt2\omega\beta)~,
\ee
where
\be
I(a)\equiv\int_0^\infty dx ~x^2\ln\left(1-e^{-\sqrt{x^2+a^2}}\right)~,
\ee
and for high temperatures leads to the Stefan-Boltzmann law 
\be
U_0\sim -\frac{\pi^2T^4}{90}~~~~\mbox{for}~~\omega\beta<<1~.
\ee
In order to achieve this, we need to take
\be\label{Rbeta}
R_\beta=\frac{1}{F_\beta(0)}\exp\left(-\frac{VI(\sqrt2\omega\beta)}{2\pi^2\beta^3}\right)~,
\ee
such that the partition function is
\bea
Z_\beta(k)&\simeq&R_\beta\Big[F_1(k)\exp(-S_1(k))+F_2(k)\exp(-S_2(k))\Big]\\
&=&R_\beta\Big[F_\beta(k)\exp(-\Sigma(k))+F_\beta(-k)\exp(-\Sigma(-k))\Big]\nn
&=&\exp\left(-\frac{VI(\sqrt2\omega\beta)}{2\pi^2\beta^3}\right)\left(2+\frac{B\omega\beta A_\beta}{32}~k^2\right)+{\cal O}(k^4)~,\nonumber
\eea
where
\bea
A_\beta&=&\frac{1}{\sinh^2(\omega\beta/\sqrt2)}\Big(-32+18\omega\beta-119B\omega\beta+(32+137B\omega\beta)\cosh(\sqrt2\omega\beta)\\
&&~~~~~~~~~~~~~~~~~~~~+3\sqrt2(7-16B\omega\beta)\sinh(\sqrt2\omega\beta)\Big)~.\nonumber
\eea
As expected the partition function is even in $k$, and one can check that $A_\beta>0$ for all $\beta$.
The classical field, the homogeneous source generated saddle point, is 
\be
\phi_c\equiv-\frac{\delta\ln(Z_\beta)}{\delta j}~~\to~~-\frac{1}{V\beta}\frac{\partial\ln(Z_\beta)}{\partial j}
=\frac{-v}{2B\omega\beta}\frac{\partial\ln(Z_\beta)}{\partial k}~,
\ee
which, in terms of the original source $j$ is
\be\label{phic1}
\phi_c=-\frac{A_\beta j}{64\omega^2}+{\cal O}(j^3)~.
\ee
Because we have a finite volume, there is a one-to-one mapping between the source $j$ and the classical field $\phi_c$,
as in the situation where quantisation is based on one saddle point.
This feature is essential to determine the Legendre transform leading to the 1PI effective potential.
Also, one can see from eq.(\ref{phic1}) that the classical field vanishes when $j\to0$, which shows that the true vacuum of the model is at $\phi_c=0$.

The effective potential is finally obtained by integrating $U_{eff}'(\phi_c)=-j$, hence
\be\label{intermTUeff}
U_{eff}(\phi_c)=U_0+\frac{32\omega^2}{A_\beta}\phi_c^2+{\cal O}(\phi_c^4)~,
\ee
where $U_0$ is given in eq.(\ref{U0}). Hence the effective potential is indeed convex, with its minimum at $\phi_0=0$,
and with volume-dependent coupling constants.

\subsection{Null Energy Condition}

For a perfect fluid with energy density $\rho$ and pressure $p$, the NEC $\rho+p\ge0$ is considered a 
universal feature for known matter \cite{NECreview}, and is one of the key ingredients for singularity theorems in Cosmology 
\cite{HP}. Few quantum effects are known for violating the NEC though \cite{NECexamples}, and among these the Casimir effect has been 
shown to provide a cosmological expansion \cite{ZS}, arising from a dynamical NEC violation. 

The competition between two vacua and the resulting tunnelling effect have proven to violate the NEC dynamically 
at zero temperature and for an $O(4)$-symmetric Euclidean spacetime \cite{AC,AP,AB}.
For the present intermediate-temperature regime though, we have $U_{eff}(0)=U_0$, 
which is independent of the volume, such that
\bea\label{rho+pintermediate}
\rho+p&=&-T\frac{\partial U_0}{\partial T}\\
&=&\frac{T^4}{2\pi^2}\left(a\frac{dI(a)}{da}-4I(a)\right)>0~,\nonumber
\eea
where $a=\sqrt2\omega\beta$: the vacuum satisfies the NEC.

\subsection{Exact SSB limit }

The results derived in this section are based on expansions which assume $B\omega\beta k\ll1$, such that in principle one cannot consider a large volume 
($B\gg1$). But it is interesting to note that eq.(\ref{intermTUeff}) remains useful for $B\gg1$ to indicate that the flatness of the effective potential, the Maxwell cut, is realized for $B=\infty$ when the SSB approximation is exact. 

One can predict a flat effective potential in the thermodynamical limit though, without any expansion of the partition function. 
Indeed, for $(B\omega\beta)^{-1}\ll k\ll1$ we have
\be
Z_\beta\to R_\beta F_\beta(-\epsilon k)\exp(-\Sigma(-\epsilon k))~~~~\mbox{where}~~~~\epsilon=\mbox{sign}(k)=\mbox{sign}(j)~.
\ee
such that, when $j\to0$,
\be
\phi_c\to\epsilon v\left(1-\frac{3}{8\sqrt2}\right)~,
\ee
leading to the discontinuity $\Delta\phi_c\simeq2v$. Hence the one-to-one mapping between $j$ and $\phi_c$ is lost for an infinite volume:
$j=0$ for all $|\phi_c|\lesssim v$, leading to a constant effective potential and thus to the Maxwell cut.

\section{Low temperature regime}
\label{SecsmallT}

For large $B\omega\beta$ but a source small enough to satisfy $B\omega\beta|k|\ll1$, the two static saddle points play a 
similar role, and the instanton/anti-instanton pairs are most of the time asymptotically close to one static saddle point or the other - see Fig.\ref{phi01}.
In the corresponding approximate saddle point, these instantons and
anti-instantons can be translated individually: they form a dilute gas when $\omega\beta$ is large enough for instanton and 
anti-instantons to be far enough from each other, and the action for an $n$-pairs configuration can be approximated by the expression (\ref{SBn1}). 
The probability per unit time for an instanton or anti-instanton to form 
is $\omega\sqrt{S_0/2}~e^{-S_0/2}$, so that the average number $\bar N$ of these configurations during the Euclidean time $\beta$ is 
\be\label{Nbar}
\bar N\simeq\omega\beta\sqrt\frac{S_0}{2}~e^{-S_0/2}~.
\ee
We note that $\bar N$ is just $\beta/\tau_t$ introduced in Section \ref{Secversus}.  
The average distance $\Delta\beta$ between instantons and anti-instantons can be expressed in terms of $\bar N$ 
\be\label{dilgas}
\Delta\beta\equiv\frac{\beta}{\bar N}\simeq\sqrt\frac{2}{S_0}~\frac{e^{S_0/2}}{\omega}~,
\ee
and is large compared to the width $1/\omega$ of an instanton or anti-instanton in the dilute gas assumption.

\subsection{Effective potential}

Taking into account the fluctuation factor (\ref{Fnbounce}),
the rescaled partition function including the static saddle points and the gas of instanton/anti-instanton pairs is
\bea\label{dilinstg}
Z_\beta(k)&=&R_\beta\Bigg[ F_\beta(k)e^{-\Sigma(k)}+F_\beta(-k)e^{-\Sigma(-k)}\\
&&+F_{\beta/2}(k)F_{\beta/2}(-k)
\left(\sum_{n=1}^\infty\frac{(\omega\beta)^{2n}}{(2n)!}\left(\frac{S_0}{2}\right)^n e^{-nS_0}\right)e^{-\Sigma(k)/2-\Sigma(-k)/2}\Bigg]\nn
&=&R_\beta\Big[F_\beta(k)e^{-\Sigma(k)}+F_\beta(-k)e^{-\Sigma(-k)}\nn
&&+F_{\beta/2}(k)F_{\beta/2}(-k)\left(\cosh(\bar N)-1\right)e^{-\Sigma(k)/2-\Sigma(-k)/2}\Big]~.\nonumber
\eea
This leads to the classical field 
\be
\phi_c=-\frac{\hat A_\beta j}{64\omega^2}+{\cal O}(j^3)~,
\ee
where, if we assume $\exp(-\omega\beta/\sqrt2)\ll1$,
\be
\hat A_\beta=32+21\sqrt2+2B\omega\beta\frac{137-48\sqrt2}{1+\cosh(\bar N)}~,
\ee
and $\bar N$ is given by eq.(\ref{Nbar}). Finally, the convex effective potential is 
\be\label{lowTUeff}
U_{eff}(\phi_c)=U_1+\frac{32\omega^2}{\hat A_\beta}\phi_c^2+{\cal O}(\phi_c^4)~,
\ee
where 
\be\label{U1}
U_1=U_0-\frac{1}{V\beta}\ln\big(\cosh(\overline{N})+1\big)~,
\ee
and $U_0$ is given by eq.(\ref{U0}).

\subsection{Violation of the Null Energy Condition}

From the expression (\ref{U1}) for the vacuum energy, one can consider two asymptotic cases:\\

\nin\underline{$\overline{N}\ll1$} This situation corresponds to the suppression of instanton/anti-instanton pairs, 
such that we expect to recover the same results as in the intermediate temperature regime. Indeed, we have
\be
U_1\simeq U_0-\frac{\ln2}{V\beta}~,
\ee
and $(T\partial/\partial T+V\partial/\partial V)(V\beta)=0$.
The sum $\rho+p$ is thus identical to the expression (\ref{rho+pintermediate}), and the NEC is satisfied.
Note that 
\be
\lim_{T\to0}(\rho+p)=0~,
\ee
which is the expected result for a zero-temperature theory in infinite volume, as long as the limits $T\to0$ and $V\to\infty$ 
are taken in such a way that $\overline{N}\ll1$.\\

\nin\underline{$1\ll\overline{N}$} In this case we have
\be
U_1\simeq U_0-\frac{\overline{N}}{V\beta}+\frac{\ln2}{V\beta}~,
\ee
such that
\be\label{rho+plow}
\rho+p=\frac{T^4}{2\pi^2}\left(a\frac{dI(a)}{da}-4I(a)\right)
-\frac{1+S_0}{2}\frac{\omega}{V}\sqrt\frac{S_0}{2}~e^{-S_0/2}~,
\ee
where $a=\sqrt2\omega\beta$. One can clearly see here the competition between:\\
{\it(i)} temperature-driven quantum corrections 
(first term), which depend on $T$ only and vanish for $T\to0$;\\
{\it(ii)} tunnelling (second term) which depends on $V$ only and vanishes for $V\to\infty$.\\
The regime $1\ll\overline{N}$ allows to fix $V$ and take $T\to0$, such that it is always possible to find
a temperature small enough for which $\rho+p<0$, and the NEC is violated in a given volume.

\section{Conclusion}

This article describes finite-temperature and finite-volume tunnelling between degenerate vacua of a scalar theory. 
Taking care of the appropriate order of the different limits to consider (large volume or/and large inverse temperature), 
we showed that the NEC can be violated dynamically by a non-extensive vacuum energy, generated by competing vacua
at any finite temperature and finite volume.

One motivation for this work is the generation of a cosmological bounce in Early Universe Cosmology \cite{AC}. 
In the vicinity of a bounce, spacetime is effectively flat, 
allowing the tunnelling process described here. Which finite volume should then be considered is still an open question: 
one might naively think of the Hubble volume, but the latter becomes infinite at the bounce, such that one needs to define another causal volume to describe tunnelling.
Furthermore, in order to apply the mechanism described here to curved spacetimes, one needs to find the energy-momentum tensor which takes into account
the non-extensive nature of the matter effective action. The first step in this direction is detailed in \cite{AP}, but it needs to be further developed in
future work.

The extension of this mechanism to curved spacetime, specifically to Cosmology, should involve a comparison between time scales for spacetime 
expansion and for tunnelling. 
Hence a study of real-time tunneling would be complementary, in order to allow a real-time-dependent process (see \cite{Ai} for a review), 
unlike the present study which is done at equilibrium. 

Finally, the energetic effects derived here could have analogue Condensed Matter systems, as those used to study false vacuum decay \cite{quantsimul}, 
which is an avenue to explore in the longer term.

\section*{Acknowledgements}

We would like to thank the Referee for critical comments on this article.
The work of JA is supported by the Leverhulme Trust (grant RPG-2021-299) and the STFC (grant ST/T000759/1).

\appendix

\section{One-loop dynamics}

\subsection{Effective theory for the order parameter}\label{wapp}

The Wilsonian action for the order parameter is a highly involved non-local functional, to be approximated here in four steps. 
First the functional form (\ref{wes}) of the local potential approximation is assumed, yielding
\be\label{olwils}
U^{(1)}(\phi)=\frac{m_B^2}2\phi^2+\frac{\lambda_B}{4!}\phi^4+\frac1{2V\beta}\mbox{Tr}\ln\Big((2\pi n\beta^{-1})^2+k^2+U''_{bare}(\phi)\Big)~
\ee
in the one-loop approximation where the trace represents the sum over Mastsubara frequency $2\pi n\beta^{-1}$ and non-vanishing three-momentum. 
Next the zero temperature local potential is truncated into a quartic form (without the source term)
\be
U^{(1)}_0(\phi)=\frac{\lambda_0}{24}(\phi^2-v_0^2)^2~,
\ee
where the index 0 denotes $T=0$,
involving the cut-off independent zero  temperature renormalised parameters. 
According to the strategy of the saddle point expansion the saddle point gives the tree-level contribution and $v={\cal O}(\lambda^0)$. 
The third approximation is to restrict the finite temperature corrections to ${\cal O}(\lambda)$, 
$m^2=m^2_0-\lambda T^2/24$ and $\lambda=\lambda_0$. This leads to action (\ref{wes}) with the known temperature-dependent vev
\be
v^2=v_0^2-\frac{1}{4\beta^2}~,
\ee
from which one can see that symmetry is restored above the temperature $2v_0$, where $v=0$. 
In this article we assume that the temperature is lower than $2v_0$ (or equivalently $\beta>\beta_c=1/2v_0$), and thermal fluctuations do not restore the symmetry.

The low energy modes are not always perturbative in the symmetry broken phase. In fact, the modes with momentum $|\vec k|<\omega=v\sqrt{\lambda/6}$ develop non-vanishing saddle point, eg. domains of the false vacuum, when the order parameter is brought in between the degenerate minimas of one-loop potential by the help of an external source. 

It is instructive to consider these approximations from the point of view of the Landua-Ginzburg double expansion in the amplitude and the derivative of the order parameter. 
In fact, our scheme corresponds to the ${\cal O}(\phi^4)$ and $\partial_t^2$ order where the wave function renormalisation constant is frozen to one. 
Higher order terms in the effective action should be taken into account when the order parameter assumes larger and faster changing value, at smaller volume, 
higher temperature and strong external source.

\subsection{Fluctuation determinant for a static saddle point}\label{1loopApp}

Starting from the action
\be
S=B\int_{-\omega\beta/2}^{\omega\beta/2}d\tau\left((\varphi')^2+\frac{1}{2}(\varphi^2-1)^2+2~k\varphi\right)~,
\ee
expressed in terms of dimensionless quantities only, one can see that $B$ plays the role of a dimensionless volume.
Also, $\omega\beta$ plays the role of a dimensionless finite total Euclidean time, imposing a discrete set of dimensionless frequencies $\nu_n=2\pi n/(\omega\beta)$.
The integration over frequencies should be replaced by summation 
\be
\int d\nu~f(\nu)~\to~\frac{1}{\omega\beta}\sum_n f(\nu_n)~,
\ee
and the Dirac distribution for frequencies becomes $\delta(\nu-\nu')~\to~\omega\beta\delta_{n,n'}$.
The trace of a time-dependent operator ${\cal O}$ is 
\bea
\mbox{Tr}\{\mathcal{O}\}&=&\int d^4x\int d^4x' \delta^{(4)}(x-x'){\cal O}_{\tau,\tau'}\\
&=&B\int d\tau\int d\tau'\delta(\tau-\tau'){\cal O}_{\tau,\tau'}~,\nonumber
\eea
which, in terms of the Fourier components $\tilde{\cal O}_{n,n'}$, gives
\be
\mbox{Tr}\{\mathcal{O}\}=\frac{B}{\omega\beta}\sum_n\sum_{n'}\delta_{n,n'}\tilde{\cal O}_{n,n'}~.
\ee
The second derivative of the action (\ref{S}) evaluated at a static saddle point $\varphi_s$ is, in terms of the discrete Fourier components, 
\be
\frac{\partial^2S}{\partial\tilde\varphi(\nu_n)\partial\tilde\varphi(\nu_{n'})}=2B\big(\nu_n^2+3\varphi_s^2-1\big)\omega\beta\delta_{n,n'}~,
\ee
and is diagonal in $n,n'$, such that its logarithm is also diagonal
\be
\ln\left(\frac{\partial^2S}{\partial\tilde\varphi(\nu_n)\partial\tilde\varphi(\nu_{n'})}\right)
=\ln\big[2B\big(\nu_n^2+3\varphi_s^2-1\big)\big]\omega\beta\delta_{n,n'}~.
\ee
The fluctuation determinant is 
\be
F(k)=\frac{1}{\sqrt{\mbox{det}(\delta^2S)}}= C\exp\left(-\frac{B}{2}\sum_{n=-\infty}^\infty\ln\big(n^2+\Phi^2(k)\big)\right)~,
\ee
where $C$ is a (source-independent) constant which can be set to 1, and
\bea
\Phi(k)=\frac{\omega\beta}{2\pi}\sqrt{3\varphi_s^2(k)-1}~.
\eea
For the sum over the modes $n$, we note that
\be
\frac{d}{d\Phi}\sum_{n=-\infty}^\infty\ln\big(n^2+\Phi^2\big)
=2\Phi\sum_{n=-\infty}^\infty\frac{1}{n^2+\Phi^2}=2\pi\coth(\pi\Phi)~,
\ee
such that
\be
\sum_{n=-\infty}^\infty\ln\big(n^2+\Phi^2\big)=2\ln|\sinh(\pi\Phi)|+~\mbox{constant}~,
\ee
where the constant does not depend on any parameter appearing in $\Phi$, and is therefore disregarded. 
The fluctuation determinant is finally
\be
F(k)=\exp\Big(-B\ln|\sinh(\pi\Phi(k))|\Big)~.
\ee

\end{document}